\documentclass[a4paper, aps, prb, reprint, superscriptaddress, twocolumn, preprintnumbers, amsmath, amssymb,nobalancelastpage, 10pt, longbibliography]{revtex4-2}
\usepackage{svg}
\usepackage[version=4]{mhchem}
\usepackage{amsmath}
\usepackage{verbatim}
\usepackage{graphicx}
\usepackage{color}
\usepackage{tikz}
\usepackage{upgreek}
\usepackage{float}
\usepackage{orcidlink}
\usepackage{textcomp}

\newcommand{\PCMO}{Pr$_{0.5}$Ca$_{1.5}$MnO$_4$}
\newcommand{\LSMO}{La$_{0.5}$Sr$_{1.5}$MnO$_4$}

\newcommand{\mjd}{mJ/cm$^{2}$}
\usepackage[strings]{underscore}

\begin{document}

\title{Orbital–lattice coupling and polaronic dressing of electronic states in \PCMO~revealed by ultrafast broadband spectroscopy}
\author{Emmanuel B. Amuah \orcidlink{https://orcid.org/0000-0003-2332-4115}}
\affiliation{ICFO - Institut de Ci\`encies Fot\`oniques, The Barcelona Institute of Science and Technology, Av. Carl Friedrich Gauss 3, 08860 Castelldefels, Barcelona, Spain}
\affiliation{Department of Physics and Astronomy, Aarhus University, Ny Munkegade 120, 8000 Aarhus C, Denmark.}

\author{Khalid M. Siddiqui \orcidlink{https://orcid.org/0000-0001-6640-2406}}
\email[]{ksiddiq1@slac.stanford.edu}
\affiliation{Department of Physics and Astronomy, Aarhus University, Ny Munkegade 120, 8000 Aarhus C, Denmark.}

\author{Maurizio Monti \orcidlink{https://orcid.org/0000-0003-2192-3747}}
\affiliation{Department of Physics and Astronomy, Aarhus University, Ny Munkegade 120, 8000 Aarhus C, Denmark.}

\author{Daniel P\'erez-Salinas \orcidlink{https://orcid.org/0000-0002-4641-8387}}
\affiliation{Department of Physics and Astronomy, Aarhus University, Ny Munkegade 120, 8000 Aarhus C, Denmark.}
\affiliation{ALBA Synchrotron Light Source, Cerdanyola del Valles, 08290 Barcelona, Spain}

\author{Hanna Strojecka \orcidlink{0009-0002-8992-3620}
}
\affiliation{Department of Physics and Astronomy, Aarhus University, Ny Munkegade 120, 8000 Aarhus C, Denmark.}
\affiliation{Univ Rennes, CNRS, Institut de Physique de Rennes – UMR 6251, 35000 Rennes, France.}

\author{Thomas H. Meyland \orcidlink{0009-0005-8344-7469}
}
\affiliation{Department of Physics and Astronomy, Aarhus University, Ny Munkegade 120, 8000 Aarhus C, Denmark.}

\author{Nikolaj Klinkby \orcidlink{0009-0008-0317-116X}
}
\affiliation{Department of Physics and Astronomy, Aarhus University, Ny Munkegade 120, 8000 Aarhus C, Denmark.}

\author{Allan S. Johnson\orcidlink{0000-0002-0711-708X}}
\affiliation{IMDEA Nanoscience, Madrid, Spain.}

\author{Simon E. Wall \orcidlink{ https://orcid.org/0000-0002-6136-0224}}
\email[]{simon.wall@phys.au.dk}
\affiliation{Department of Physics and Astronomy, Aarhus University, Ny Munkegade 120, 8000 Aarhus C, Denmark.}

\date{\today}

\begin{abstract}
The interplay between long-range structural distortions and orbital order, and local short-range phenomena such as polarons, are key for understanding the properties of the manganites. In this work, we examine the role of local and long-range physics in relation to orbital ordering in the single-layered manganite \PCMO~ with a combination of optical reflection anisotropy and ultrafast broadband pump-probe spectroscopy. We find that the reflection anisotropy, measured in equilibrium, is strongly sensitive to charge and orbital-ordering transition only. However, the ultrafast response, measuring the non-equilibrium state, is sensitive to a range of phenomena in the material, including delocalized (long wavelength) phonons as well as localized polarons. In particular, we find that a strong sensitivity to electronic and structural changes give rise to apparent nonlinearities when probed at specific wavelengths, despite the key degrees of freedom remaining linear. These observations point to the important role of orbital–lattice couplings and polarons dressing the electronic states across the \PCMO~phase diagram.    

\end{abstract}

\maketitle

\section{Introduction}
Structural and electronic ordering are ubiquitous in strongly correlated materials, often leading to broken symmetries, and give rise to diverse and emergent functionalities~\cite{Dagotto2005,daSilvaNeto-2014,Tokura2017}. Understanding how these ordering phenomena manifest is key to controlling the complex behaviors of correlated systems. To this end, mixed-valence manganites have provided a fertile platform for studying the intricate interplay between electronic, structural, orbital, and spin degrees of freedom that govern material properties~\cite{Mitchell2001,Tomioka2009}, hosting different long-ranged ordered states of charges, spin and orbitals. The three-dimensional A$_{1-x}$A$^\prime_x$MnO$_3$ compounds have been studied in great detail due to their rich variety of ordered phases, colossal magnetoresistance effects~\cite{Jin_1994,Dagotto_2001,Chahara1993}, and light-induced phase transitions~\cite{Fiebig1998,Polli2007,Rini2007,Rini2009,liFemtosecondSwitchingMagnetism2013,Wall_2012,Ichikawa_2011}.

The single-layered Ruddlesden-Popper manganites, A$_{1-x}$A$^\prime_x$MnO$_4$, have a suppressed colossal magnetoresistance effect in comparison to their three-dimensional counterparts~\cite{Tokunaga_1999}, but are often regarded as model systems because the layered structure means that the physics of interest is confined to the two-dimensional manganese oxide planes \cite{RonnowRennerAeppliKimuraTokuraNat2006}. \LSMO~(LSMO) in particular has played a critical role in understanding orbital ordering~\cite{Dhesi_2004,Wilkins_2003,Castleton2000} and its dynamics following optical~\cite{Ogasawara2001,Ehrke2011a,Tobey2012,perez-salinasMultimodeExcitationDrives2022a,montiDisentanglingHeterogeneityDisorder2024}, mid-IR~\cite{tobeyUltrafastElectronicPhase2008,Forst2011} and THz excitation~\cite{Miller_2015}. However, other single-layered compounds have received less attention, particularly when driven out-of-equilibrium despite showing much richer phase diagrams. Understanding the response in these materials can help elucidate the roles of lattice and electronic correlations in determining the material properties and how to control them. \PCMO~(PCMO) is such a material. 

\begin{figure}[t]
    \centering
    \includegraphics[width=8.6 cm]{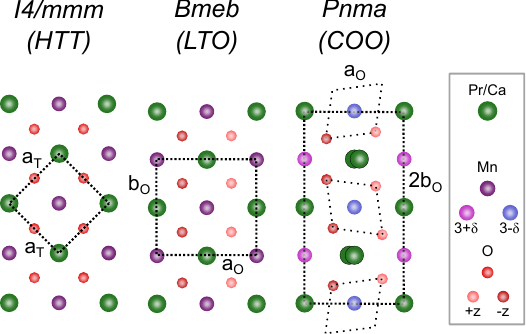}
    \caption{Structure of \PCMO~in the $ab$ plane in each phase based on reported diffraction data~\cite{Yu2007,porerCorrelationsElectronicOrder2020,Capogrosso2010thesis,Schumann2010thesis}. In the $Bmeb$ phase the oxygen ions rotate around the $a$ axis resulting in a displacement along the out-of-plane $z$-direction. In the $Pnma$ phase, the Jahn-Teller distortion doubles the periodicity along the $b$ axis resulting in the Mn ion being situated in two inequivalent positions, which results in a small charge disproportionation, $\delta$. HTT: High-temperature tetragonal ($T> 466$\,K); LTO: Low-temperature orthorhombic ($320<T< 466$\,K); COO: charge and orbital ordering ($T< 320$\,K).}
    \label{fig:structure}
\end{figure}

Orbital ordering in LSMO emerges from a tetragonal ($I4/mmm$) phase on cooling below $T_c \sim 220$\,K and below $T_N = 110$\,K antiferromagnetic order sets in. PCMO, while similar, has a richer phase diagram. The current understanding of the crystal structure of PCMO comes from a combination of electron diffraction and X-ray diffraction (XRD) studies~\cite{Yu2007, porerCorrelationsElectronicOrder2020,Capogrosso2010thesis,Schumann2010thesis}, with the reported structures depicted in Fig.~\ref{fig:structure}. Like LSMO, at high temperatures, PCMO is in a high-temperature tetragonal phase (HTT) with equal \textit{a} and \textit{b} axes lengths, and is characterized by the space group $I4/mmm$. However, already below $T_{LTO} = 466$\,K, the system enters a low-temperature orthorhombic phase (LTO) due to an out-of-plane rotation of the oxygen octahedra that breaks the in-plane C$_4$ symmetry. In addition, strain develops along the $b$ axis, disrupting the equivalence between the \textit{a} and \textit{b} axes, which results in an orthorhombic unit cell represented by the $Bmeb$ space group. At $T_\text{COO}\sim320$ K, the system undergoes a charge and orbital ordering (COO) transition, accompanied by a significant increase in resistivity and enhanced in-plane symmetry breaking concomitant with Jahn-Teller (JT) distortion of the crystal. Within this phase, the charges form a checkerboard pattern, often represented with alternating Mn$^{3+}$ and Mn$^{4+}$ sites, although the actual charge disproportionation, $\delta$, is typically very small \cite{herrero-2011}. Concurrently, the $e_g$ electron orbitals, $3y^2-r^2$ and $3x^2-y^2$ on \ce{Mn^{3+}} sites align in a zigzag pattern along the crystallographic $b$ axis. This transition doubles the length of the $b$  axis and the structure is characterized by the $Pnma$ space group, although some ambiguity remains about this assignment~\cite{Schumann2010thesis}. 

Relevant to this may be the existence of small polarons: a patchy network of short-coherence length, disordered and localized lattice distortions, which have been discussed in the literature based on Raman and electron diffraction studies of layered manganites \cite{Romero1998,MartinCarron2002,Li1999SmallPolaron,Wu2007ZenerPolaron}. These structural modulations have been attributed to the distortion of \ce{Mn^{4+}} octahedra, and have been used to explain thermally activated transport \cite{Kressdorf2020HotPolaron} and other effects such as magnetoresistance in manganese perovskites \cite{Millis1998LatticeEffects}. At lower temperatures, further structural transitions occur in PCMO and at $T^* = 146$\,K, the system moves into a low-temperature-less orthorhombic (LTLO) phase where the difference between \textit{a} and \textit{b} lengths (denoted $a_\text{O}$ and $b_\text{O}$ in Fig.~\ref{fig:structure}) shrinks, which is believed to arise from a change in the oxygen tilting scheme~\cite{porerCorrelationsElectronicOrder2020}. Finally, 3D antiferromagnetism sets in at $T_N = 130$ K, with two-dimensional magnetic fluctuations possibly extending up to 200 K~\cite{chiEffectAntiferromagneticSpin2007}. 

The ultrafast structural response to optical excitation in PCMO was previously studied by Porer \textit{et al.} using time-resolved XRD at 100 K, where the amplitude of the Bragg peak associated with COO was tracked~\cite{porerCorrelationsElectronicOrder2020}. The COO phase was found to melt in less than 100 fs and the strong role of lattice was reported by the authors. While the limited time resolution of the measurement meant that the Jahn-Teller mode could not be observed directly~\cite{Polli2007,Wall2009, Singla2013, Matsuzaki_2009}, the melting and recovery of the COO phase was still suggested to be tightly coupled to it, thus demonstrating the key connection between structural degree of freedom and long-range order. However, being a structural probe, direct information about the electronic changes was not available, which motivates further investigation of the interplay of the electronic order with other degrees of freedom following optical excitation on ultrafast timescales, including in other phases of PCMO. Furthermore, the recent observations of the ultrafast formation of polarons in LSMO~\cite{montiDisentanglingHeterogeneityDisorder2024} suggests that local physics should also play a role in PCMO. 

Here, we examine the optical and ultrafast response of PCMO using a combination of angle-resolved birefringence measurements for equilibrium symmetry characterization, and ultrafast broadband spectroscopy for probing the dynamics of the electronic states and coherent lattice modes on a timescale that can directly measure the Jahn-Teller mode. Our work reveals the existence of local order in the low-temperature orthorhombic phase of PCMO\textemdash which we attribute to structural disorder and dynamical polaronic fluctuations\textemdash and a pronounced rotation of the optical axis upon entering the COO phase. The latter observation is at odds with the currently assigned space group of the JT-distorted charge/orbital phase \cite{Schumann2010thesis}. Moreover, the ultrafast response across different phases reveals intriguing trends, and analysis of the coherent phonon response presented here uncovers a breakdown in the linear relationship between the induced phonon displacement and the measured changes in reflectivity at specific probe wavelengths. Together, these observations showcase the complex and rich nature of the ultrafast response in PCMO, which we discuss in the context of electronic structure changes shaped by orbital order and polaronic effects. 

\section{Methods}
Single crystalline samples of PCMO were grown by floating zone technique \cite{blascoGrowthSr23Ln12008}, polished to give a (001) surface normal and mounted in an optical cryostat.  Temperature-dependent optical anisotropy measurements were performed using the setup described in Ref.~\cite{siddiquiVersatileSetupSymmetryresolved2025} with a probe wavelength of 1300\,nm and white light continuum. Briefly, the probe light was sent through a combination of a polarizer (which fixed the incident and final detected polarizations) and rotating half-wave plate which controls the polarization state of the light impinging on the sample. The reflected light was retro-transported through the same rotating half-waveplate, thereby undoing the applied rotation, leaving only the polarization rotation imparted by the anisotropic sample. The light then passed through a polarizer, which converted the polarization rotation into an intensity modulation which was detected by a photodiode in the case of the NIR probe or a spectrometer for the white light continuum. The intensity was recorded as a function of half-wave plate angle, $\theta$ which corresponds to a polarization angle $\Phi$ = -2$\theta$. 

High time-resolution pump-probe experiments were performed with 1.78 \textmu m pump derived from the idler output of a NIR optical parametric amplifier (OPA) as described in Ref.~\cite{amuahAchromaticPumpProbe2021a}. The light was spectrally broadened in a hollow-core fiber and subsequently compressed to give a duration of 15 fs full-width-at-half-maximum (FWHM). The visible probe, on the other hand, was produced by a commercial non-collinear OPA emitting broadband light centered at 630 nm and of duration 15 fs. Pump and probe polarizations were parallel and along \textit{a}/\textit{b} axis of the crystal. The probe pulse was dispersed inside a home-built, prism-based spectrometer and detected by a calibrated CCD array on a shot-to-shot basis. The repetition rates were set at 1 kHz (probe) and 500 Hz (pump), and the differential reflectivity as a function of probe wavelength was calculated as the difference between the pumped and unpumped reflectivity at each time delay, $\frac{\Delta R(\lambda, t)}{R_0}$. Each probe spectrum was normalized by a corresponding reference white light spectrum, $R_0$ to account for any light intensity fluctuations. The raw data for all measurements are available in Ref.~\cite{amuah2025-Zenado}.

\section{Results}
\subsection{Optical anisotropy}
\begin{figure}
    \centering
    \includegraphics[width=8.6 cm]{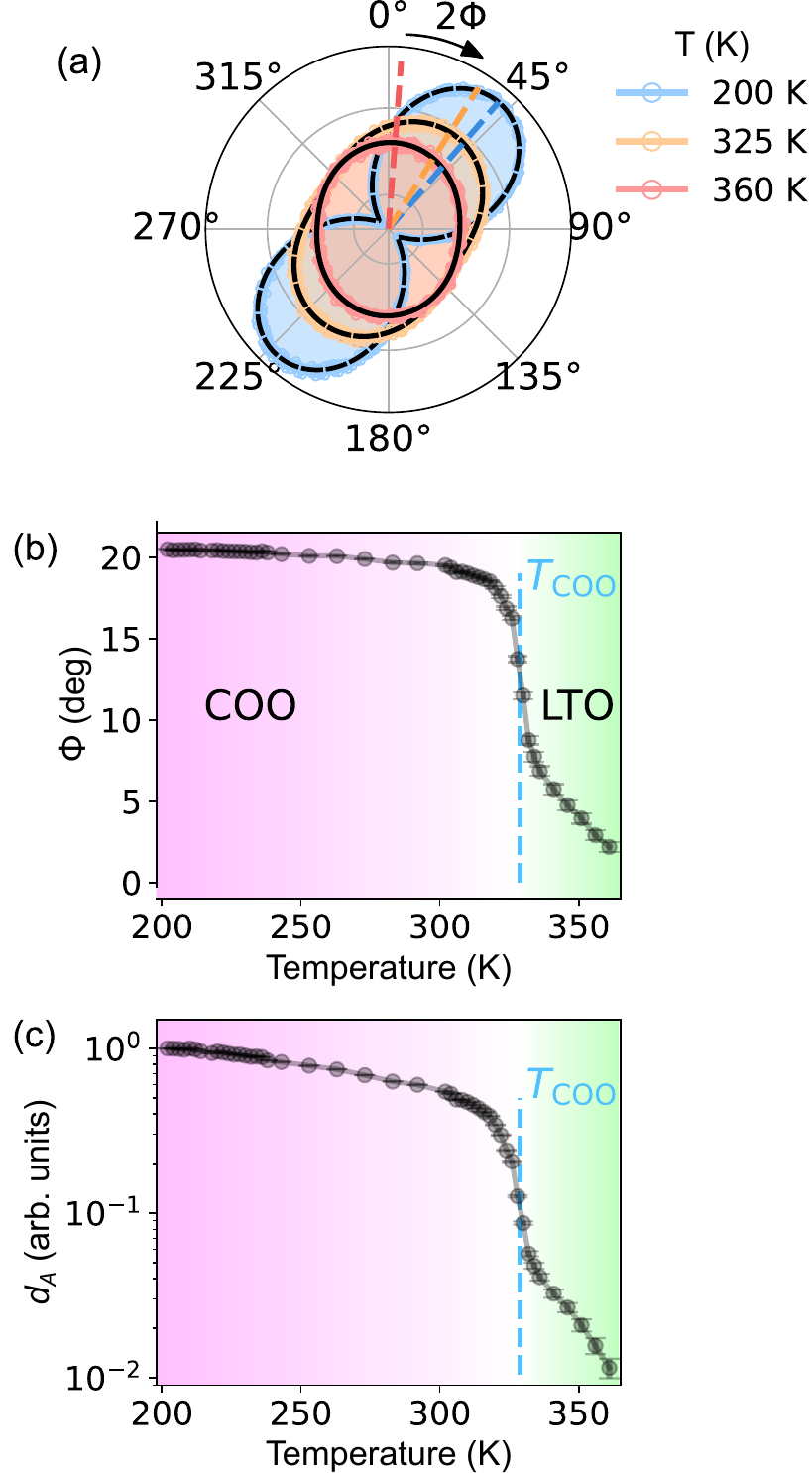}
    \caption{(a) Polarization-resolved reflectivity of \PCMO~at three temperatures. The $Bmeb$ phase at $T = 200$ K (blue) in the charge/orbital ordered, at the phase transition temperature $T = 325$ K (orange), and in the LTO phase at $T = 360$ K (red). Markers are measurements and black dashed lines are fits to the data using the model described in the main text. $\Phi = 0$ corresponds to the probe light along the $a$ crystallographic axis in the $Bmeb$ phase. Colored dashed lines indicate the orientation of the optical axis relative to the crystallographic axis. (b) Temperature-dependent rotation of the optical axis as a function of temperature. Blue dashed vertical line marks the transition temperature, $T_\text{COO}$ and the shading separates COO phase from LTO. (c) Extracted order parameter $d_A$ as a function of temperature on a semi-log scale. Above $T_{\text{COO}}$, a small anisotropy is still present, potentially indicating some short range order persists above $T_{\text{COO}}$. Error bars reflect one standard deviation uncertainty computed from several identical measurements.}
    \label{fig:bi}
\end{figure}

We start by examining the temperature-dependent reflection anisotropy to characterize the orbital ordering phase transitions in PCMO. Figure~\ref{fig:bi}a shows the polarization-dependent reflectivity of PCMO in the charge/orbital phase (blue, T = 200 K) as well as at (orange, T = 325), and above (blue, T = 360 K) the COO phase transition temperature, represented in polar form as a function of polarization angle, $\Phi$. Anisotropy is observed at all three temperatures reflecting the fact that, unlike LSMO, PCMO remains orthorhombic above $T_{\text{COO}}$. The LTO phase shows only a weak anisotropy despite the orthorhombic structure, but the anisotropy becomes significantly larger when the system enters the charge/orbital ordering phase. Curiously, we observe that the optical axis (indicated by the dashed lines) also rotates significantly between the two phases. 

Figure~\ref{fig:bi}b shows the extracted angle which the optical axis makes relative to the crystallographic axis as a function of temperature. When cooling towards $T_\text{COO}$, the pattern rotates by several degrees before jumping by over 20$^\circ$ at $T_\text{COO}\sim 325$ K. Although not shown here, the optical anisotropy showed no further changes when cooled below $T^*$.  This is surprising given that the system becomes less orthorhombic in LTLO phase, which may be expected to reduce the birefringent signal. This could be related to formation of a multi-domain state, with domains smaller than the probe spot size, or points to the fact that the charge/orbital ordering is by far the strongest perturbation to the electronic structure at this probed wavelength, consistent with recent experiments on thin film samples which have shown that charge/orbital ordering is insensitive to the small oxygen rotations found in the LTLO phase~\cite{flathmannRelationshipStructureCharge2024}.

This large change in the orientation of the optical axis is unexpected. The polarization rotation observed in PCMO is significantly larger than the approximately 3$^\circ$ observed in LSMO over the orbital-ordering phase transition~\cite{DaniThesis2022}. Previous studies have measured the reflectivity at fixed polarizations (parallel and perpendicular with respect to the $Bmeb$ phase crystallographic axes) and thus, could not sperate rotations of the anisotropy from changes in the magnitude~\cite{Singla2013, leeInPlaneAnisotropyElectronic2006,majidiTemperaturedependentAnisotropicOptical2013}. In addition, the crystal structure determined by diffraction for both LTO and COO phases indicates that the optical axis should lie along the $\vec{a}$ direction in each phase. If this were true then the onset of charge/orbital ordering should not be expected to induce a rotation of the optical axis, contrary to what is observed. 

While strain induced by mounting the sample could add an additional symmetry breaking along an arbitrary direction, and thereby cause a rotation of the optical axis upon the onset of orbital ordering, the required mounting strain would need to be significantly larger than that already induced by the LTO phase. The latter already breaks the symmetry between the $\vec{a}$ and $ \vec{b}$ directions by several percent\cite{ZafarPetkovAbeykoonJPhysMater2024,MarisVolotchaevPalstraNJP2004}, and therefore, it seems unlikely that mounting strain can contribute significantly to large rotation in the optical axis as observed here.  

As there is some ambiguity about the assignment of the $Pnma$~\cite{Schumann2010thesis}, we suggest that an alternative space group may be needed for the low temperature state. To explain this rotation, we suggest that the two sources of birefringence in the material, the strain associated with the LTO phase and the orbital order, induce anisotropies along different directions (see Appendix~\ref{appendix:A}). We let the LTO-phase crystallographic $\vec{a}$ and $\vec{b}$ axes define our coordinate system and then assume that orbital order induces a symmetry along the $\vec{a}+\vec{b}$ direction, which corresponds to $B_{1g}$ symmetry perturbation in the point group of the LTO phase. The optical anisotropy can, therefore, be expressed as, 
\begin{equation}    r = 
    \begin{pmatrix}
        r_a& 0 \\
        0& r_b
    \end{pmatrix}
    + \frac{d(T)}{2}
    \begin{pmatrix}
        0& 1 \\
        1& 0
    \end{pmatrix},
    \label{eq:rmain}
\end{equation}

where $r_{a/b}$ are the reflectivity coefficients for the electric field polarized along the $a/b$ axes and $d(T)$ is the temperature-dependent symmetry-breaking order parameter. In principle, all parameters are complex and $r_{a/b}$ may also be temperature dependent. Using this model, we fit our data to extract a quantity $d_A$ that is proportional to $d$ see Appendix~\ref{appendix:A} for details. 

Figure~\ref{fig:bi}c displays the temperature dependence of $d_A$. It can be seen that there is finite symmetry-breaking above $T_\text{COO}$, which is more than an order of magnitude weaker than when the system crosses into the charge/orbital ordered phase. This observation points to the persistence of some local short-range order (e.g. small COO domains or polarons~\cite{Romero1998}) well above $T_\text{COO}$. 
\begin{figure}
    \centering
    \includegraphics[width=7.5 cm]{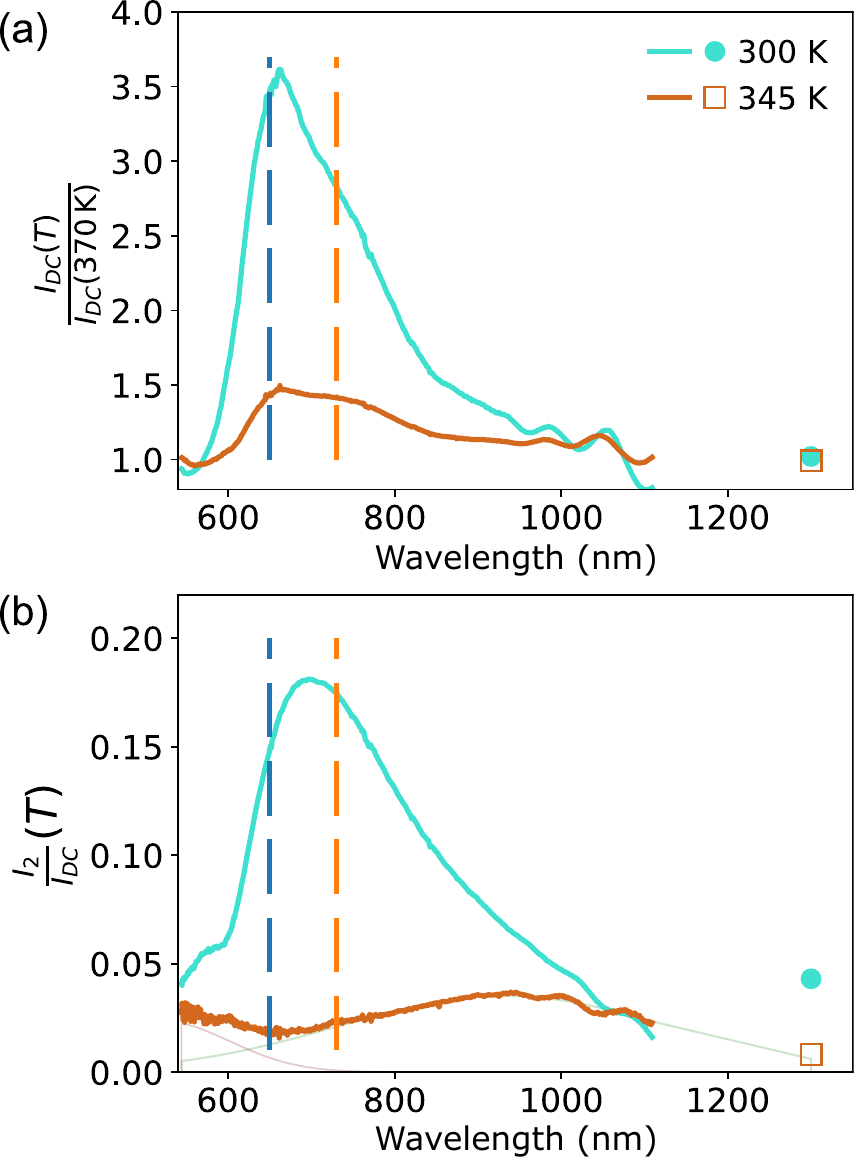} 
    \caption{Wavelength-dependent changes in isotropic (a) and anisotropic (b) reflectivity at $T=300$\,K (teal line and solid circle) and $T=345$\,K (dark brown line and open square). (a) The isotropic reflectivity ($I_{DC}(T)$), is normalized to the value at 370 K, i.e. $I_{DC}(T)/I_{DC}(370K)$. (b) The anisotropic reflectivity, $I_2$, is normalized by the corresponding $I_{DC}$ term at the same temperature. Blue and orange dashed vertical line denote 650 nm and 730 nm probe wavelengths used to analyze the broadband pump probe data later in the text. The light drown and green lines in (b) are guides to the eye showing the emergence of two broad resonances. }
    \label{fig:3}
\end{figure}
Additionally, we repeated the anisotropy measurements using a broadband white light probe spanning the visible range. Fig.~\ref{fig:3}a shows the the isotropic, or angularly averaged, reflectivity, $I_{DC}$ above (345\,K) and below (300\,K) $T_{\mathrm{COO}}$ relative to the reflectivity at 370\,K. While the isotropic reflectivity at 1300\,nm is relatively independent of temperature, a large increase is found between 600-800\,nm consisting of a double peak structure. This feature can still be seen without a significant shift in the resonance position above $T_{\mathrm{COO}}$, but the amplitude is greatly reduced. 

Fig.~\ref{fig:3}b shows the magnitude of the anisotropic reflectivity, $I_2$, normalized to $I_{DC}$ for the two temperatures in (a). The anisotropy term is directly connected to the strength of the orbital ordering (see Appendix~\ref{appendix:A}). A strong resonance is observed at similar range to that seen in the isotropic reflectivities for temperatures below $T_{\mathrm{COO}}$. However, above $T_{\mathrm{COO}}$, the profile splits into two features, one at higher photon energies outside our probed regions and the other in the near IR. This will have implications for the ensuing discussion on the ultrafast dynamics of PCMO.

\subsection{Ultrafast perturbative response }
\begin{figure}[t]
    \centering
    \includegraphics[width=8.4 cm]{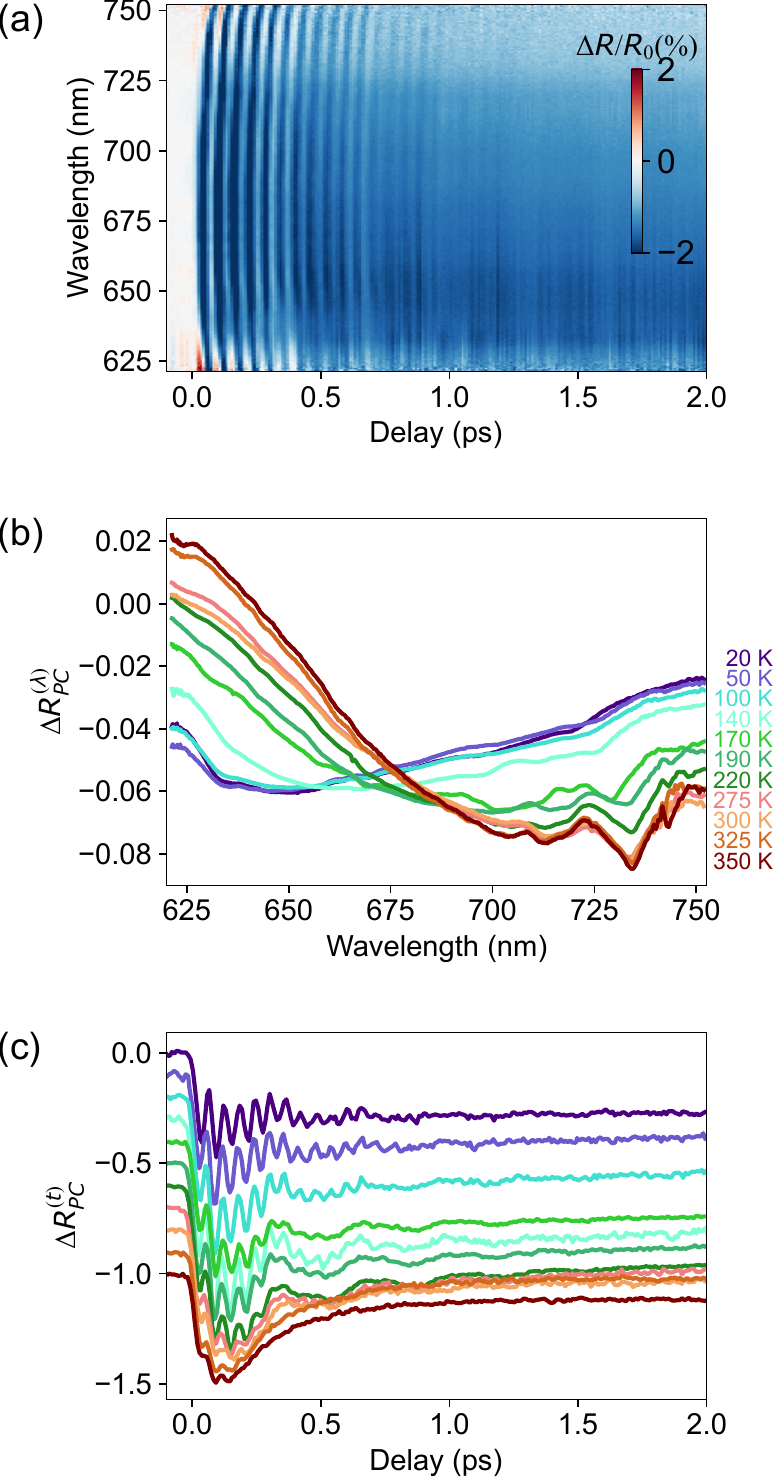}
    \caption{Transient dynamics of the broadband reflectivity at 2.5 mJ/ cm$^2$. (a) Typical broadband change in reflectivity measured at 20 K showing a decrease with oscillations due to coherent phonons modulating the transient signal. (b) Spectral dependence of the dominant principal component obtained from the two-dimensional reflectivity data set for all measured temperatures. (c) Corresponding temporal dynamics extracted from the PCA analysis.} 
    \label{fig:lowPC}
\end{figure}

Next, we study the non-equilibrium dynamics of PCMO after being subjected to weak optical excitation at 2.5~\mjd~in order to characterize the linear response of each phase with a high time resolution and to gain insight into the electronic and the structural response through the coherent phonon spectra. Figure~\ref{fig:lowPC} shows typical transients measured by our setup. Fig.~\ref{fig:lowPC}(a) depicts the broadband response measured at 20 K showing ultrafast suppression in transient reflectivity that is additionally modulated by fast oscillations across all probe wavelengths. To analyze the temperature-dependent data, we first perform a principal component analysis (PCA) that decomposes each dataset into a set of significant principal components that capture the variance in signal as a function of wavelength and time (see Appendix~\ref{appendix:B}). While more than one component is needed to fully describe the transient response~\cite{amuahDeterminationCorrectionSpectral2024}, we find that the dominant component gives a good representation of the key dynamics in each phase. 

Figure~\ref{fig:lowPC}b displays the spectral response at each recorded temperature, while the temporal response derived from the dominant principal component is shown in Fig.~\ref{fig:lowPC}c. Note that both components are plotted as negative to reflect the fact that the signal measured in Fig.~\ref{fig:lowPC}a is predominately negative. The full reconstruction of the data in Fig.~\ref{fig:lowPC}a is then given by 
\begin{equation}
    \frac{\Delta R}{R}(\lambda, t) = -\Delta R^{(t)}_{PC} \otimes \Delta R^{(\lambda)}_{PC}.
\end{equation}

The temperature dependence of the two principal components is discussed in the next sections.

\subsubsection{Spectral dependence}
    
\begin{figure}
    \centering
    \includegraphics[width=8.6 cm]{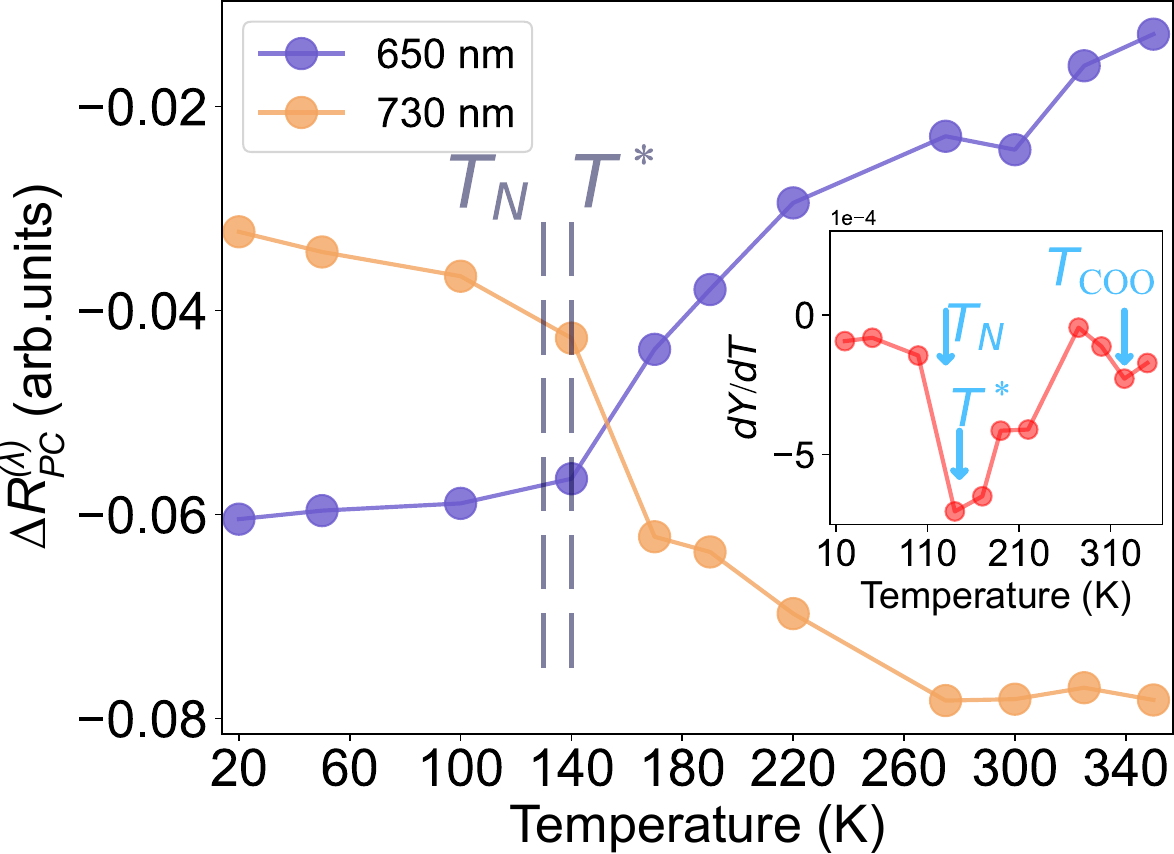}
    \caption{Temperature-dependent transient reflectivity changes at selected colors. The change is strongest at $T^*$. The inset shows the derivative of $Y$, which is the difference in the change of reflectivity between 730 and 650 nm, with respect to temperature $T$. The slope exhibits the largest sign change at $T^*$.}
    \label{fig:lowPCWaveDiff}
\end{figure}

We start by discussing the temperature dependence of the spectral response (Fig.~\ref{fig:lowPC}b). At low temperatures, the magnitude of the change is strongest at shorter wavelengths, and the entire measured window shows a reduction in reflectivity. The change is roughly temperature-independent when heating, up to $T^*$, at which point the changes at shorter wavelengths are suppressed, and the dominant response now appears at longer wavelengths. This shift in spectral weight continues until $T_{\text{COO}}$, after which the changes become more or less temperature-independent at long wavelengths, and a small positive signal becomes apparent at shorter wavelengths. To more clearly show this redistribution of spectral weight, Fig.~\ref{fig:lowPCWaveDiff} illustrates how the change in reflectivity at bluer (650\,nm) and redder (730\,nm) measured wavelengths evolves against temperature, revealing opposite trends: the 650 nm signal increases upon heating, while the 730\,nm signal decreases. Computing the derivative of $Y = \Delta R_{730~\mathrm{nm}} - \Delta R_{650~\mathrm{nm}}$ with respect to temperature $T$, as shown in the inset of  Fig.~\ref{fig:lowPCWaveDiff}, highlights that the largest change in slope with a sign reversal occurs at $T^*$. The significant shift in spectral shape at $T^*$, is in contrast to the optical birefringence, which showed no sensitivity in this region, confirming that $T^*$ has a significant effect on the unoccupied states of PCMO. The sensitivity of the differential reflectivity at $T^*$, instead of $T_N$, suggests these changes result from the structural transition associated to $T^*$, rather than magnetic effects which can also modulate the pump-induced reflectivity change~\cite{Pastor_2022}. However, due to the proximity of $T^*$ and $T_N$, the effects of magnetic ordering on the electronic structure cannot be excluded. 

\subsubsection{Coherent phonon dynamics}
\begin{figure}
    \centering
    \includegraphics[width=8.6 cm]{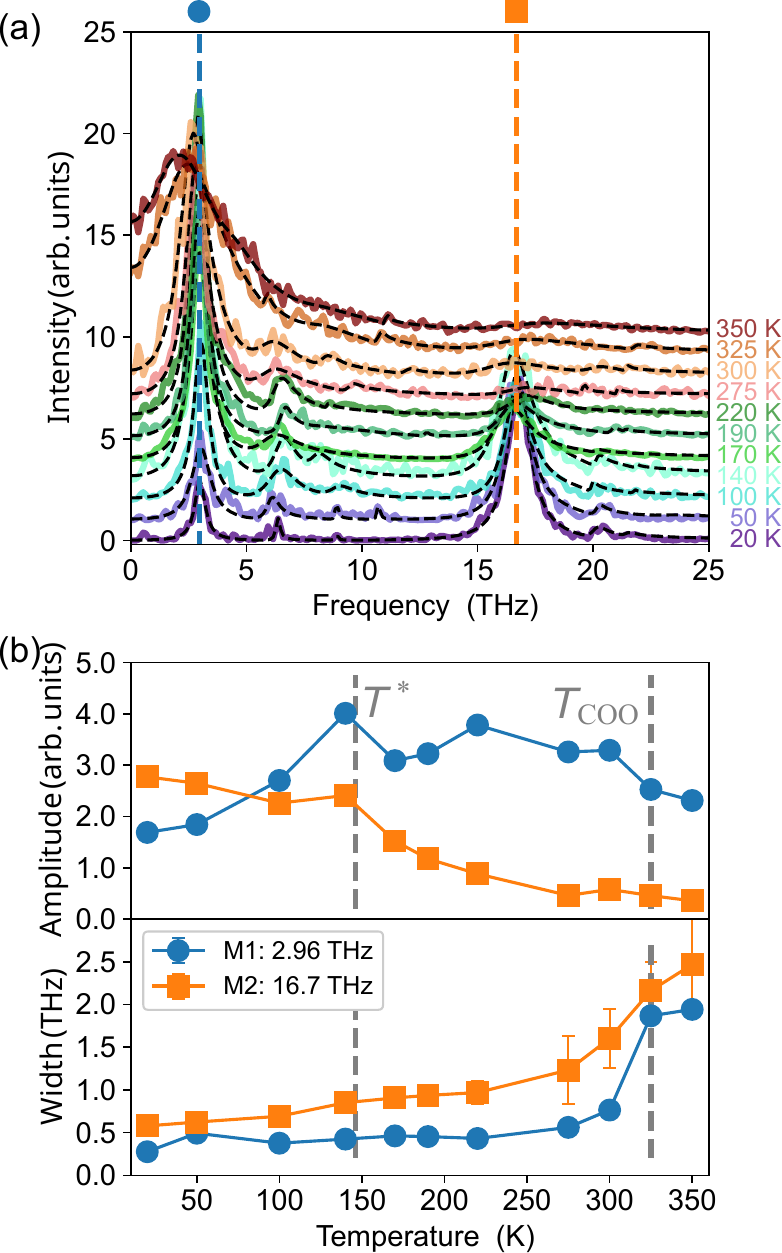} 
    \caption{Coherent phonon response as a function of temperature. (a) Temperature-dependent phonon spectra obtained by Fourier transformation of oscillations observed in Fig.~\ref{fig:lowPC}b following electronic background removal. Plots are vertically displaced for clarity. Dashed black lines are fits using Eq.~\ref{Eq:FFTfit}. Two of the phonon modes are examined in the text, and are labeled by square and circle markers. (b) Intensities of the two selected modes against temperature (upper panel, note M2 mode in orange is multiplied by 2 for easier comparison with M1) and their widths (lower panel). Error bars represent uncertainties at one standard deviation obtained from the fits and subsequently error propagated.}.
    \label{fig:lowPhonon}
\end{figure}

We now turn our attention to the coherent dynamics that are imprinted in the temporal response. To isolate the phonon contribution from the non-oscillatory dynamics, we fit the dominant PCA component shown in Fig.~\ref{fig:lowPC}b using the procedure outlined in the Appendix~\ref{appendix:B}.
\begin{figure*}[ht]
    \centering
    \includegraphics[width=15.5 cm]{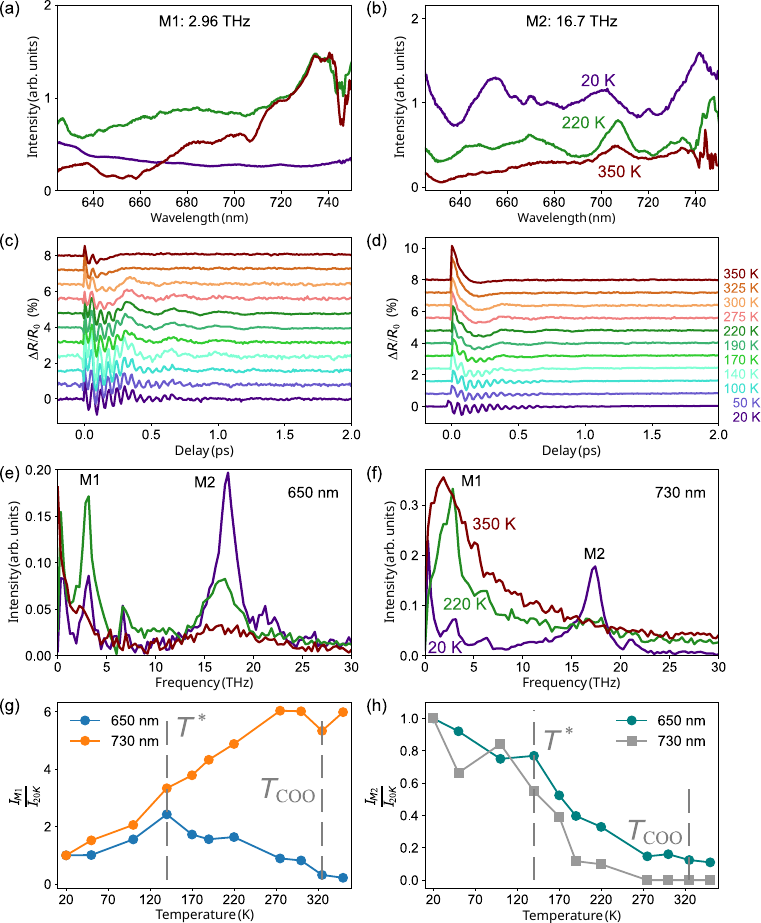} 
    \caption{Temperature and wavelength-dependent amplitudes of phonon modes (a) M1 and (b) M2. Shaded regions indicate the spectral ranges used for integration. (c/d) Time-dependence from shaded regions in (a/b) corresponding to integration centered at 650 nm/730 nm. (e/f) Power spectra at three different temperatures measured at 650 nm and 730 nm. (g/h) Normalized intensities of the FFT peak signal for the M1/M2 modes at 650 and 750 nm after fitting with the complex Lorentzian oscillator model discussed in Appendix~\ref{appendix:B}. The vertical dashed line mark the $T^*$ and $T_\text{COO}$ transitions. }
    \label{fig:lowPhononWL}
\end{figure*}
Figure~\ref{fig:lowPhonon} provides an overview of the temperature dependence of the coherent phonon signal when pumped with 2.5~\mjd. Four distinct modes are well-resolved at 2.96, 6.4, 16.7 and 20.1 THz (99, 213, 557 and 670 cm$^{-1}$). With the exception of the low frequency, 2.96 THz mode, all other modes were observed in previous equilibrium Raman measurements which showed that they are of $A_g$ symmetry~\cite{Handayani2015}. These four modes also appeared in pump-probe measurements of a related compound, LSMO~\cite{perez-salinasMultimodeExcitationDrives2022a}, but all modes are systematically higher in frequency in the PCMO compound. The lowest frequency mode shows the strongest shift, increasing by almost 10\% from 2.7 THz in LSMO. This is consistent with the mode being assigned to the A-site ionic motion, which are almost a factor of two lighter in PCMO~\cite{amelitchevStructuralChemicalAnalysis2001}. The 6.4 THz mode has previously been attributed to tilting or bending of the \ce{MnO6} octahedra around the $b$ axis. The 16.7 THz mode corresponds to the Jahn–Teller–active distortion (asymmetric stretch) of the \ce{Mn^{3+}} sites, while the 20 THz mode is associated with symmetric \ce{M-O} stretching of the \ce{Mn^{4+}} octahedra \cite{Handayani2015}. These phonon modes are Raman-active in the LTO and COO phases.

All modes can be observed in all three phases to varying degrees, but become highly damped above $T_{\text{COO}}$. Their presence can be attributed to the small fraction of localized, disordered regions (i.e. polaron-orbital disorder \cite{Li1999SmallPolaron,Wu2007ZenerPolaron}) persisting well into the LTO phase, in line with the observations made with optical anisotropy measurements in Fig.~\ref{fig:bi}c. In the following, we focus on the two most prominent modes: the low-frequency 2.96 THz mode (M1) and the high-frequency 16.7 THz mode (M2), extracting their widths and amplitudes by simultaneously fitting all peaks at each temperature as described in Appendix~\ref{appendix:B}.

As can be seen from Fig.~\ref{fig:lowPhonon}b, the M2 mode, has the highest amplitude at low temperatures. On heating, the amplitude of this mode slowly decreases until $T^*$, at which point the rate of change increases, similar to what is observed in the non-oscillatory dynamics shown in Fig.~\ref{fig:lowPCWaveDiff}, and the amplitude plateaus close to $T_{\text{COO}}$, in good agreement with Raman scattering~\cite{Handayani2015}. The damping rate, which is proportional to the phonon peak width, shown in the lower panel of Fig.~\ref{fig:lowPhonon}b, displays a small moderate increase at $T^*$, but rapidly increases at $T_{\text{COO}}$. Unlike the cubic compound~\cite{Wall2009}, no significant change in damping due to magnetic ordering is observed, suggesting that magneto-elastic coupling is not as strong in the doped single-layered compound, consistent with the reduced colossal magnetoresistance effect in the layered compounds.

The temperature dependence of the M1 peak is more unusual. The amplitude increases with increasing temperature to $T^*$ where it becomes temperature independent until $T_{\text{COO}}$, after which it begins to fall again. In contrast to the M2 mode, the damping rate of the M1 mode is almost temperature independent up to $T_{\text{COO}}$. Above this temperature, there is a significant broadening. We note that, in general, in many systems \textemdash including correlated quantum materials, the amplitude of the coherent phonon signal remains constant or decreases with increasing temperature \cite{zhaiCoherentPhonons2024} therefore, the increase in the amplitude of the M1 mode is unexpected and unusual. 

To better understand the atypical response of the M1 mode compared to M2, we examined how the phonon amplitudes vary with the probe wavelength more accurately. To do this, we exploit the full data set, rather than the principal component considered thus far (see Appendix~\ref{appendix:B} for details). Figures~\ref{fig:lowPhononWL}(a, b) show the wavelength-dependent, area integrated FFT signals of M1 and M2 modes, respectively at three different temperatures. At 20 K, the M1 mode is relatively independent of wavelength, with a slightly greater sensitivity at shorter wavelengths. Similarly, while the M2 mode shows more structure, the response is quite wavelength independent. However, at higher temperatures, the spectral sensitivity of the M1 mode changes significantly: the mode sensitivity becomes significantly stronger at longer wavelengths and suppressed at shorter, similar to the spectral response of the incoherent background, $\Delta R^{\lambda}_{PC}$ shown in Fig.~\ref{fig:lowPC}b. The M2 mode, on the other hand, shows a suppression at all wavelengths.

To address this more quantitatively, we explore two regions centered at 650 and 730\,nm and 20\,nm wide. The time traces for these two regions are shown in Figs.~\ref{fig:lowPhononWL}(c-d) for all temperatures. It is clear that each wavelength region has a different sensitivity to the M1 and M2 modes. The reflectivity at 650\,nm (Fig.~\ref{fig:lowPhononWL}(c)) is strongly modified by the M2 mode, and multiple cycles can still be clearly seen above $T_{\mathrm{COO}}$. Conversely, the M2 mode is barely visible above $T_{\mathrm{COO}}$ in the 730\,nm data (Fig.~\ref{fig:lowPhononWL}(d)), but a near single cycle of the M1 mode can still be seen. The phonon spectra at 650 nm and 730 nm following wavelength-dependent fitting of the data is captured in Figs.~\ref{fig:lowPhononWL}(e-f).

Figs.~\ref{fig:lowPhononWL}(g,h) plot the intensity of the FFT of this data for the M1 and M2 modes, respectively, as a function of temperature for the two wavelength regions considered. Now it is clear that the unusual temperature dependence of the M1 mode displayed in Fig.~\ref{fig:lowPhonon}b is, in parts, due to a diverging behavior as a function of probe wavelength. This appears to affect the M1 mode significantly more than M2. 

These results show a breakdown in the relationship between the measured phonon amplitude and the actual atomic displacements. The response of the M1 mode at 730\,nm suggests that the phonon amplitude increases with temperature up to $T_{\mathrm{COO}}$, whereas the response at 650\,nm suggests that the phonon amplitude only increases until $T^*$ before decreasing. Before discussing the origin of this effect in more detail, we compare how the phonon response changes as a function of the pump fluence rather than temperature. 

\subsection{Strong excitation below $T^*$}

\begin{figure}
    \centering
    \includegraphics[width=7.6 cm]{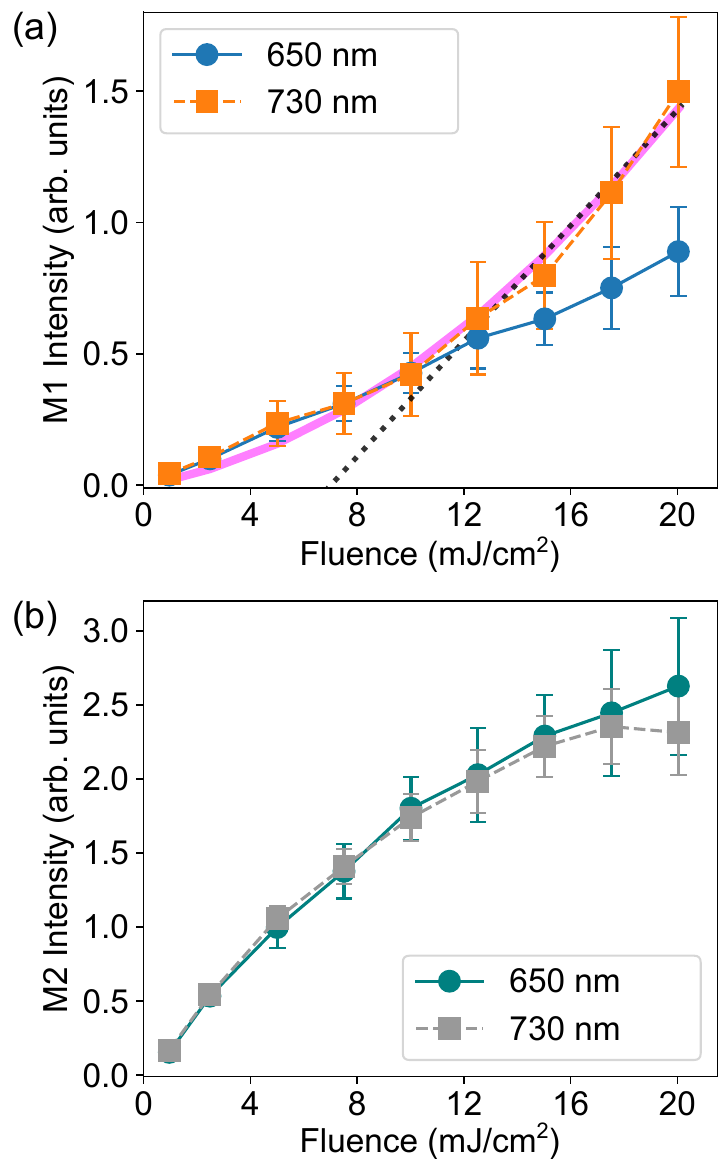}
    \caption{Fluence dependence of the (a) M1 mode and (b) M2 mode at two distinct wavelengths of 650 nm and 730 nm at 20 K. While M2 mode shows a sublinear behavior leading to saturation at higher fluences for both wavelengths, the M1 mode amplitude exhibits different scaling depending on the wavelength. The thick pink lines are fit using Eq.~\ref{eq:nonlin}. }
    \label{fig:fluence20}
\end{figure}

The observed variations in the phonon amplitude with wavelength raise an interesting question, namely, can we relate the changes in the amplitude of the coherent phonon signal in reflectivity to actual changes in the amplitudes of underlying atomic positions? Typically, the amplitude of the phonon response is taken as a direct measure of the amplitude of the actual atomic displacements, up to a scale factor. However, our data casts doubts about this assumption. To test the possible implications of this, we perform a fluence dependence of the signal at 20 K. 

In Fig.~\ref{fig:fluence20}, we plot the fluence dependence of the M1 and M2 phonon mode amplitudes, at 650 nm and 730 nm, as before. As with temperature, the amplitude of the M1 as a function of fluence depends strongly on the probe color. The response at 650\,nm is linear over the measured range, but the signal at 730\,nm shows a non-linear increase. Whereas the M2 mode shows a saturating response, with a much weaker dependence on the probe wavelength, as observed with temperature changes. 

\section{Discussion}
The results of our investigation reveal that the choice of probe wavelength significantly impacts the temperature and fluence dependence of the observables in the system when driven out of equilibrium. Specifically, we are able to detect signatures of the phase transition at $T^\star$ in the spectral dependence of the reflectivity change (Fig.~\ref{fig:lowPCWaveDiff}) and observe contrasting temperature and fluence dependence for the phonon modes, particularly the M1 mode above $T^\star$. In the following, we will discuss how the complexity in the spectral behavior can prove to be an advantage for understanding the strong interplay between structural distortions in the manganites and their effects on the electronic properties of the material. 

The static optical conductivity ($\sigma_1$) of PCMO has been measured by Majidi \textit{et al.}~\cite{majidiTemperaturedependentAnisotropicOptical2013}. They observed two broad features in the optical conductivity that overlap in the energy range measured here: a broad feature at $\sim$0.8\,eV, which was assigned to the formation of a Jahn-Teller gap, and which blue-shifts on cooling, and a second feature at $\sim3.5$\,eV, which corresponds to a charge transfer process from the oxygen $2p$ to manganese $3d$ states. The measured optical anisotropy was strongest at the Jahn-Teller gap for the orbital ordering transition, whereas the charge transfer gap was only sensitive to the anisotropy below $T^*$. These assignments are similar to Kressdorf \textit{et al.} \cite{Kressdorf2020HotPolaron} who attributed a broad absorption feature at 0.8\,eV to polarons, a feature at 1.3\,eV to on-site transition between the Jahn-Teller split $e_g$ levels and a feature at 3.5\,eV to the charge transfer excitations. In general, similar features are also seen in the three dimensional manganites. The presence of these resonances also strongly affects the equilibrium Raman response when the probe wavelength is changed, with modes associated with the Mn-O bond strongly enhanced at 2.5\,eV~\cite{Dediu2000}. Therefore, in order to understand the unusual wavelength dependence of the coherent phonon response, we must also understand how photoexcitation changes the electronic structure.

Below $T^\star$, photoexcitation has the strongest effect at shorter wavelengths (higher photon energy), while above $T^\star$ the effect is reversed and long wavelengths (low energies) are primarily affected (Fig~\ref{fig:lowPC}b). This shift originates from changes in the electronic structure of the unoccupied states associated with the charge transfer resonance~\cite{majidiTemperaturedependentAnisotropicOptical2013} due to the structural distortions that occur at $T^*$~\cite{porerCorrelationsElectronicOrder2020}. Below $T^*$, photoexcitation is primally modulating the charge transfer states, that appear below $T^*$, which we detect through the tail of the broad resonance at 3.5\, eV. Above $T^*$, these states are no longer accessible, which suppresses the signal at shorter wavelengths. Instead, the strongest changes occur at longer wavelengths, which we attribute to the ability to form polarons, which we justify based on the coherent phonon response.

We start by considering the anomalous temperature dependence of the M1 mode, which shows an initial growth in amplitude below $T^\star$, before diverging behavior sets in depending on the probe wavelength. In conventional Raman scattering, anti-stokes scattering strongly depends on the thermal population of the mode, and thus the temperature. By contrast, in a stimulated Raman process, most materials show very little change with temperature when far from a phase transition \cite{zhaiCoherentPhonons2024}. Temperature-dependent changes in the mode amplitude could arise from changes in lattice anharmonicity or a change in absorption coefficient of the pump light. We can rule out both factors. The anharmonicity is directly measured in our experiments via the damping rate or the width of the phonon peak. In order for the peak height to increase, the anharmonicity would need to reduce and the peaks should narrow with increasing temperature, which is not observed (See Fig.~\ref{fig:lowPhonon}b). Changes in the absorption coefficient can similarly be excluded as this would result in a simultaneous increase in M1 and M2 modes, which is not observed. This indicates that the effect results mainly from changes in the electronic structure measured by the probe rather than from the phonon generation mechanism.

From a symmetry perspective, the M1 and M2 modes should only appear below $T_{\mathrm{COO}}$ as they become Raman active when the system breaks translation symmetry and adopts orbital ordering. Therefore, it should be expected that the amplitude of the modes should decrease as $T_{\mathrm{COO}}$ is approached from below. This is the case for the M2 mode and also M1 when probed at 650\,nm. However, the amplitude of the M1 continues to increase up to $T_{\mathrm{COO}}$ and then saturates when probed at 730\,nm. In reality, equilibrium Raman scattering the M1 and M2 modes are never fully suppressed above  $T_{\mathrm{COO}}$ because disorder can relax the Raman selection rules. This can originate from the intrinsic A-site disorder due to Pr/Ca size difference which naturally produces such local distortions at \ce{MnO6} octahedra. This type of local structural disorder is ubiquitous in 3D and layered manganites \cite{RodriguezMartinezAttfieldPRB54R15622, ZafarPetkovAbeykoonJPhysMater2024}. Alternatively, thermal disorder and polarons can also contribute to this effect~\cite{Dediu2000, MartinCarron2002}. 

The above implies that 730\,nm  is probing a polaronic state, or localized phonon mode, whereas 650\,nm is more sensitive to the long-range zone center phonon mode, which is why it is strongly suppressed at $T^*$. A similar process has been proposed for \ce{YBa2Cu3O7} \cite{NovelliFausti2017}. This is consistent with our broadband anisotropy data shown in Fig.~\ref{fig:3} which shows that the residual birefringence observed above $T_{\mathrm{COO}}$ (Fig~\ref{fig:bi}) is associated with broad resonance in the NIR region, which enhances the ability to observe the polarons, and the spectral shape of light induced changes that are observed above $T^*$.

While the dependence of the M2 mode is less unusual, there are still some puzzling factors which we can now explain with a combination of changes in the resonance conditions and polaron formation. The M2 mode is attributed to Jahn-Teller oscillations of the oxygen ions and is strongly associated with charge and orbital ordering. Yet, the amplitude of this mode does not track the temperature dependence of the orbital order, but instead correlates with $T^*$ at all probe colors. As we have discussed, $T^*$ has a strong effect on the charge transfer resonance and the charge transfer resonance affects the visibility of the Jahn-Teller modes~\cite{Dediu2000}. Therefore, when the charge transfer resonance is modified at $T^*$, the ability to detect the phonon is strongly reduced. Above $T_{\mathrm{COO}}$, the presence of disorder and polarons will also enable excitation of the oxygen modes, but we only have sufficient sensitivity when probing the charge transfer resonance.

Having shown that resonance shifts and polarons play a significant role in understanding the temperature-dependent data, we now use this knowledge to interpret the fluence data. The pump induced change in reflectivity can then be expanded in a power series of the electronic and phononic degrees of freedom to first order as~\cite{Zeiger1992},

\begin{equation}
    dR(\lambda, t)  = \sum_i \frac{dR}{dQ_i}(\lambda)Q_i(t) + \frac{dR}{dT_e}(\lambda) T_e(t)+...,\label{eq:lin}
\end{equation}

where $Q_i$ is the amplitude of the $i^\text{th}$ mode and $T_e$ is the electronic temperature, neither of which depend on wavelength. The wavelength dependence comes from the pre-factors, which can be due to changes in the positions of the bands driven by coherent phonons, or changes in the matrix elements due to a change in electronic distribution \cite{Moskvin2010,Yang2019}. We choose to model the electronic response with an effective temperature, even though the reality is likely more complex, as we found the pump induced dynamics were very similar, up to a scale factor, when pumping with 1.78\,\textmu m (0.69\,eV) and 800\,nm (1.55\,eV). This shows that the initially occupied states do not play a major role in the dynamics and the photoexcited populations is rapidly scattered into a distribution independent of the pumping wavelength, just as was observed in \LSMO~\cite{perez-salinasMultimodeExcitationDrives2022a, montiDisentanglingHeterogeneityDisorder2024}

Equation~\ref{eq:lin} describes reflectivity changes that are linearly proportional to the phonon amplitude or the electronic temperature. The system can respond non-linearly with fluence if either the electronic $T_e$ or phonon amplitudes $Q_i$ change non-linearly with fluence, or if the reflectivity depends non-linearly on the phonon or electronic amplitude, while the underlying degree of freedom still responds linearly.  

In the former scenario, Eq.~\ref{eq:lin} remains valid and the non-linearity should be seen equally at all probe colors where the linear response was seen. In the latter scenario, Eq.~\ref{eq:lin} should be expanded to second-order and $Q_i$ and $T_e$ remain linear. Expanding to second order results in two additional terms that can contribute to an oscillating signal, one proportional $Q^2$, and a cross term between the electronic and phonon degrees of freedom. The $Q^2$ term will lead to phonon harmonics, which are not seen here and must, therefore, be weak. Therefore, we expand to second order keeping the cross term only so that the change in reflectivity can be written as 

\begin{equation}
    dR_Q(\lambda, t)  = \left( \frac{dR}{dQ}(\lambda) + \frac{d^2R}{dQdT_e}(\lambda) T_e(t) \right)Q(t), \label{eq:nonlin}
\end{equation}

Now, if both the electronic temperature and the phonon mode amplitude are linearly dependent on fluence, the detected phonon response can be non-linear. As $\frac{dR}{dQ}(\lambda)$ and $\frac{d^2R}{dQdT_e}(\lambda)$ are two different functions of wavelength, this non-linearity will depend on the choice of probe color consistent with the observations in Fig.~\ref{fig:fluence20}. This has important consequences for how coherent phonon data are interpreted. 

For example, Singla \textit{et al.}~\cite{Singla2013} extrapolated the fluence dependence of the Jahn-Teller mode and found a threshold-like behavior for generating the phonon mode. A similar situation appears when considering only the high fluence data measured at 730 nm for the M1 mode (dashed line in Fig.~\ref{fig:fluence20}). However, the phonon amplitude is linear and shows no sign of a threshold if the probe wavelength is set to 650\,nm. As no threshold was observed here for the M2 mode, or in other studies of the M2 mode in LSMO~\cite{perez-salinasMultimodeExcitationDrives2022a}, we suggest that the apparent threshold observed by Singla and co-workers was a result of a second order contribution to the reflectivity. In addition, Ning \textit{et al.} observed certain modes saturating as a function of pump fluence, while others remained linear in a ruthenate sample and suggested that this was an indication of a light-induced phase transition \cite{ning2023}. This is similar to what is seen when we compare the M1 and M2 modes at 650\,nm. Our work suggests that such a response can also arise from the second-order response of the reflectivity while the underlying physical property, the phonon amplitude, remains linear.

\section{Conclusions}
In summary, our combined optical birefringence and ultrafast spectroscopy experiments reveal new insights into the complex phase diagram of PCMO. Upon entering the COO phase, the optical axis undergoes a pronounced rotation, indicative of a phase with potentially lower symmetry than the currently proposed $Pnma$. Moreover, evidence of local short-range order of the COO phase was observed above $T_\text{COO}$, attributed to small polarons and JT fluctuations which break local symmetry and enable the appearance of symmetry-forbidden phonons above the phase transition temperature.

Among the phonon modes, the 2.96 THz mode, linked to A-site ionic motions, exhibits a strong wavelength and temperature-dependence, including displaying sub- and superlinear scaling with fluence depending on the probe wavelengths caused by polaron excitation and orbital-lattice couplings. This coupling results in a breakdown in the linear relationship between the amplitude of the detected phonon-induced change in reflectivity and the actual phonon amplitude. 

Finally, the complex behaviors uncovered here, stemming from rich light-matter interactions, distinguish PCMO from other layered manganites like LSMO, making it more susceptible to external stimuli and highly amenable to photoengineering thereby offering a platform to explore new, possibly hidden, phases.

\begin{acknowledgments}

We thank Javier Blasco for providing the sample and Urs Staub for insightful comments on the structure. We acknowledge funding from the Novo Nordisk Foundation (NNF23OC0084990) and support from the Center for Integrated Materials Research (iMAT) at Aarhus University. A.S.J. acknowledges the support of the Ramón y Cajal Program (Grant No. RYC2021-032392-I). IMDEA Nanociencia acknowledges support from the “Severo Ochoa” Programme for Centers of Excellence in R\&D (MICIN, CEX2020-001039-S).
\end{acknowledgments}

\section*{Data Availability Statement}
All data used in this manuscript is available from a Zenodo repository~\cite{amuah2025-Zenado}.

\appendix

\section{Optical Anisotropy}
\label{appendix:A}
\setcounter{equation}{0}
\renewcommand{\theequation}{A.\arabic{equation}}
The in-plane reflectivity of PCMO can be represented as a $2\times 2$ matrix, which can be decomposed into an isotropic contribution $A_0$ and an symmetry breaking anisotropic contribution ${\bm \Delta}$
\begin{equation}
    {\bm r} = A_0 {\bm I} + {\bm \Delta}\sigma_z
\end{equation}

where ${\bm I}$ is the identity matrix and ${\bm \sigma}_z$ is the Pauli $\sigma_z$ matrix. In the absence of strain, ${\bm \Delta}=0$ in the high-temperature, $I4/mmm$, phase. 

Symmetry breaking produces a difference in reflectivity along two orthogonal directions which can be expressed as 
\begin{equation}
    {\bm \Delta} = 2\Delta {\bm R}^T(\phi){\bm \sigma}_z{\bm R}(\phi),
\end{equation}

where $\Delta$ is the magnitude of the difference between the reflectivities along the two different directions, $\phi$ is the angle between one axis (the axis with the higher reflectivity) and the laboratory frame $x$-axis, ${\bm R}(\phi)$ is the 2D rotation matrix. This can be expanded as 
\begin{equation}
    {\bm \Delta} = 2\Delta\left( \cos (2 \phi){\bm \sigma}_z - \sin (2 \phi){\bm \sigma}_x\right),\label{eq:delta_g}
\end{equation}

where ${\bm \sigma}_x$ is the Pauli $\sigma_x$ matrix.

The anisotropic reflectivity is probed by the angular dependence of porlarization of the probe light, as described in~\cite{siddiquiVersatileSetupSymmetryresolved2025}. 
The probe light polarization is rotated in the $ab$-plane of the crystal, making an angle $\theta$ to the $x$-axis of the laboratory frame 
\begin{equation}
    E_i = E_0=
    \begin{pmatrix}
    \cos\theta \\ 
    -\sin\theta \\ 
    \end{pmatrix}
\end{equation}

The polarization component of the reflected light parallel to the incident light is detected. The field in this configuration can then be expressed as  

\begin{equation}
\begin{aligned}
    E(\theta) &=\frac{1}{E_0} E_i^T r E_i \\
    &= E_0 A + 2 E_0 \Delta \left( \cos(2\phi) \cos(2\theta)- \sin(2\phi) \sin(2\theta) \right) \\
    &=E_0 A + 2 E_0 \Delta \cos(2\theta+2\phi),
\end{aligned} 
\end{equation}

The detected intensity is then 
\begin{equation}
\begin{aligned}
    I_r(\theta) = &I_i(|A|^2+2 |\Delta|^2) \\
    &+ 2 I_i(\Delta^* A+\Delta A^*)\cos(2\theta+2\phi) \\
    &+ 2 I_i |\Delta|^2\cos(4\theta+4\phi),    
\end{aligned}
\end{equation}

where $I_i/r$ corresponds to the incident/reflected intensity of the probe light. 

Finally, we fit the anisotropic reflectivity with the following fit function

\begin{equation}
I_r(\theta) = I_{DC} + I_2\cos(2\theta+2\phi_2) + I_4\cos(4\theta+4\phi_4),
\label{eq:Ifit}
\end{equation} 

for each temperature. Although we should expect $\phi_2=\phi_4=\phi$ from the above equation, we will see below that the complex nature of the refractive index can add an additional complication which breaks this relation. 

In PCMO, ${\bm \Delta}$ has two contributions to the anisotropy: the strain associated with the transition to LTO phase, which defines the $a$ and $b$ axes of the LTO phase,${\bm \Delta}_{LTO}$ and the contribution from orbital ordering, which should only appear below $T_{\mathrm{OO}}$, ${\bm \Delta}_{\mathrm{OO}}$. Thus, the total anisotropy can be written as  

\begin{equation}
{\bm \Delta} = {\bm \Delta}_{LTO} + {\bm \Delta}_{oo},
\label{eq:delta_pcmo}
\end{equation} 

We assume that the $ab$ axes are aligned along the $xy$ axes of the lab frame. As the crystal is fixed, this form of anisotropy cannot rotate as a function of temperature, so, 

\begin{equation}
{\bm \Delta}_{LTO} = 2 \alpha(T) {\bm \sigma}_{z}.
\end{equation} 

With this definition, the reflectivity along the $a/b$ axis of the LTO phase is given as $r_{a/b}= A \pm 2\alpha$ (see  Eq.~\ref{eq:rmain} of the main text. 

The orbital ordering contribution also cannot rotate as a function of temperature. We assume that it introduces an asymmetry at 45 degrees to the $a$ axis. This introduces an anisotropy of the form  

\begin{equation}
{\bm \Delta}_{\mathrm{OO}} = 2 d(T) {\bm \sigma}_{x},
\end{equation} 

which corresponds to a perturbation of the point group of $B_{1g}$ symmetry.  

If $\alpha$ and $d$ are real quantities, then \ref{eq:delta_pcmo} can be written in the form of \ref{eq:delta_g} as, 

\begin{equation}
{\bm \Delta} = 2 \sqrt{\alpha^2 +  d^2}\left(\frac{\alpha}{\sqrt{\alpha^2 +  d^2}}{\bm \sigma}_{z} + \frac{d}{\sqrt{\alpha^2 +  d^2}}{\bm \sigma}_{x}\right).
\end{equation} 

From this we see that $\Delta \equiv \sqrt{\alpha^2 +  d^2}$ and $\cos(2\phi) \equiv \frac{\alpha}{\sqrt{\alpha^2 +  d^2}}$. The temperature dependence of the angle of the anisotropy pattern then reflects the changing ratio of the strength of $\alpha$ to $d(T)$.

However, $\alpha$ and $d$ are related to the dielectric function and can be complex quantities, which adds an additional complexity. When taking this into account, the final equation for the angular dependent reflectivity can be written in the form of equation~\ref{eq:Ifit} with 
\begin{align}
I_{DC}/I_i &= |A|^2 +2 |\alpha|^2 + 2 |d|^2, \\ 
I_{2}/I_i &= \sqrt{\alpha_A^2 + d_A^2 }\label{eq:I2},\\
I_{4}/I_i &= \sqrt{\delta_{\alpha d}^2 + d_\alpha^2 }    
\end{align}

Where, $\alpha_A =  4\Re\{A^*\alpha\}$, $d_A =  4\Re\{A^*d\}$, $\delta_{\alpha d}=2(|\alpha|^2-|d^2|)$ and $d_{\alpha}=2\Re\{\alpha^*d\}$. 

In addition, we have

\begin{align}
\cos 2\phi_2 &= \frac{\alpha_A}{\sqrt{\alpha_A^2 + d_A^2 }} \label{eq:cos2term}\\
\cos 4\phi_4 &= \frac{\delta_{\alpha d}}{\sqrt{\delta_{\alpha d}^2 + d_\alpha^2 }}
\end{align}

We can extract $d_A$ and $\alpha_A$ using equations~\ref{eq:I2} and~\ref{eq:cos2term}. Although we do not measure $I_i$ directly, we know that it should be larger than $I_{DC}$. We assume that the isotropic reflectivity at the probe wavelength is 25\% but the general trend remains unchanged for different values. 

\section{PCA and transient fitting}
\label{appendix:B}

In this section, we briefly describe the use of principal component analysis (PCA) to aid the fitting of the electronic background and to
provide an overview of the multidimensional data set \cite{amuahDeterminationCorrectionSpectral2024}

We define the transient reflectivity matrix, \textit{M}, 
\begin{equation}
M (\lambda, t) = \frac{\Delta R (\lambda,t)}{R(\lambda)},
\end{equation}

where each column corresponds to a probe wavelength $\lambda$ and each row corresponds to a pump–probe delay $t$. PCA aims to identify orthogonal directions of maximal variance in the data. For a mean-centered dataset, the covariance matrix, $C$ is given as,

\begin{equation}
C  = \frac{1}{m-1}M^{T}M,
\end{equation}

where $m$ corresponds to the number of time delays. Diagonalization of $C$ yields a set of row and column eigenvectors, which define the principal components in wavelength and time space respectively, and their corresponding eigenvalues, which quantify the variance captured by each component. Consequently, the dataset can be expressed as a sum of separable temporal and spectral contributions
\begin{equation}
M(\lambda, t)  = \sum_i T_i(t)\otimes \Lambda_i(\lambda), 
\end{equation}

where $T_i (t)$ are orthogonal temporal components capturing the dominant dynamical processes, and $\Lambda_i(\lambda)$ are the corresponding spectral responses that quantify how strongly each temporal component modulates the reflectivity at each probe wavelength, and $\otimes$ is the outer product.

The majority of the data can be captured by a single principal component (PC0), but some finer details are missing. The data presented in Figs.~\ref{fig:lowPC}, \ref{fig:lowPCWaveDiff} and \ref{fig:lowPhonon} are based on analyzing only the time dependence of the PC0 component or the wavelength dependence. 

To extract the phonon signal, we first need to remove the slowly varying background. To do that we  using the following model function which works for all principal components, 

\begin{equation}
\begin{aligned}
T^{(i)}_{f} = \Theta(t-t_0)\bigg[ &
A_1 e^{-(t-t_0)/t_1} + A_2 \\
& + \sum_{i=1}^{2} B_i e^{-(t-t_0)/\tau_{f i}}
\cos\!\left(2\pi f_i (t-t_0)\right)
\bigg]
\label{Eq:ExponentialModel}
\end{aligned}
\end{equation}

where $\Theta$ is the Heaviside step function and $t_0$ denotes time zero. The parameters $A_1$ and $t_1$ represent the amplitude and decay time of the electronic relaxation, respectively, while $A_2$ accounts for the long-time offset. The oscillatory components have amplitudes $B_i$, with corresponding frequencies $f_i$ and damping times $\tau_{f i}$. $t_0$ needs to be fitted for each scan as slow drifts in the lab environment resulted in a change of $t_0$ by 50 fs over the two day measurement window. In addition, we also need to include the two dominant phonon modes (M1 and M2) in the fit in order to accurately obtain $t_0$. 

To extract the non-oscillatory background, the oscillatory term is disregarded (i.e $B_i$ is set to zero while the other fit parameters remain fixed), yielding a smooth exponential response. This background is subtracted from the the fitted principal component, and the resulting residuals are Fourier-transformed to obtain the phonon power spectra at each temperature. In the case of Fig.~\ref{fig:lowPhonon}, this is the the residual of the principal time component ($\Lambda_0$) minus the fit. 

However, the phonon response is observed in several principal time components. To extract the full wavelength we noted that the slowly varying background has only significant amplitude in the first three principal components. 

We then construct the background subtracted transient reflectivity matrix, $\Delta M$, by 
\begin{equation}
    \Delta M(\lambda, t) = \sum_{i=0}^2 (T_i(t)-T^{(i)}_{f})\otimes \Lambda_i(\lambda) + \sum_{i>2} T_i(t)\otimes \Lambda_i(\lambda), 
\end{equation}

This matrix is the Fourier transformed for each wavelength to give a 2D matrix of phonon frequency and optical probe wavelength. 

To fit the Fourier transforms we model each oscillator as a complex Lorentzian of the form,
\begin{equation}
F_p(\omega) = A_p\left( \frac{1}{\Gamma_p-i(\omega+\omega_p)}+ \frac{1}{\Gamma_p-i(\omega-\omega_p)} \right)
\label{Eq:FFTfitAmp}
\end{equation}

where $A_p$ corresponds to the oscillator strength of phonon $p$, $\omega_p$ is the phonon frequency, and $\Gamma_p$ is related to damping and hence, peak width. 

The data is then fitted with 

\begin{equation}
I_{FT} = |\sum_p F_p(\omega)|^2 
\label{Eq:FFTfit}
\end{equation}

From this model, the phonon intensities are extracted as $(\frac{A_p}{ \Gamma_p})^2$

\bibliography{PCMO}

\end{document}